\documentclass[conference]{IEEEtran}
\usepackage{cite}
\usepackage{soul}
\usepackage{amsmath,amssymb,amsfonts}
\usepackage{algorithmic}
\usepackage{graphicx}
\usepackage{textcomp}
\usepackage{xcolor}
\usepackage[center]{caption}
\usepackage{subfig}
\usepackage{multirow}
\usepackage{graphicx}
\usepackage[abs]{overpic}
\usepackage{pifont}

\def\BibTeX{{\rm B\kern-.05em{\sc i\kern-.025em b}\kern-.08em
    T\kern-.1667em\lower.7ex\hbox{E}\kern-.125emX}}
\begin{document}

\title{
Atomic-Scale Insights into the Switching Mechanisms of RRAM Devices

}
\makeatletter 
\newcommand{\linebreakand}{%
  \end{@IEEEauthorhalign}
  \hfill\mbox{}\par
  \mbox{}\hfill\begin{@IEEEauthorhalign}
}
\makeatother 
\author{

\IEEEauthorblockN{Md Tawsif Rahman Chowdhury}
\IEEEauthorblockA{\textit{Electrical and Computer Engineering} \\
\textit{Wayne State University}\\
Detroit, USA\\
mtawsifrc@wayne.edu}
\and
\IEEEauthorblockN{Alireza Moazzeni}
\IEEEauthorblockA{\textit{Electrical and Computer Engineering} \\
\textit{Wayne State University}\\
Detroit, USA\\
alireza@wayne.edu}
\and
\IEEEauthorblockN{Gozde Tutuncuoglu}
\IEEEauthorblockA{\textit{Electrical and Computer Engineering} \\
\textit{Wayne State University}\\
Detroit, USA\\
gozde@wayne.edu}
%
}
\maketitle

\begin{abstract}
The growing energy demands of information and communication technologies, driven by data-intensive computing and the von Neumann bottleneck, underscore the need for energy-efficient alternatives. Resistive random-access memory (RRAM) devices have emerged as promising candidates for beyond von Neumann computing paradigms, such as neuromorphic computing, offering voltage-history-dependent switching that mimics synaptic and neural behaviors. Atomic-scale mechanisms, such as defect-driven filament formation and ionic transport, govern these switching processes. In this work, we present a comprehensive characterization of Tantalum Oxide based RRAM devices featuring both oxygen-rich and oxygen-deficient switching layers. We analyze the dominant conduction mechanisms underpinning resistive switching and systematically evaluate how oxygen stoichiometry influences device behavior. Leveraging a bottom-up design methodology, we link material composition to electrical performance metrics—such as endurance, cycle-to-cycle variability, and multilevel resistance states—providing actionable guidelines for optimizing RRAM architectures for energy-efficient memory and computing applications.


\end{abstract}


\section{Introduction}


The rapid escalation in global computing demand is pushing traditional computing architectures toward their energy limits. In this context, the energy required to transfer data between memory and processing units—commonly known as the von Neumann bottleneck \cite{1}—has emerged as the primary constraint, often surpassing the energy consumed by the computations themselves \cite{2}.
This challenge has prompted extensive research into novel computing architectures and device technologies designed to overcome the inherent limitations of conventional systems. Specific efforts focus on developing energy-efficient paradigms that minimize data movement and tightly integrate memory and computation. In-memory computing has appeared as a promising alternative for data-intensive computations, as it overcomes the von Neumann bottleneck by co-locating data and computation, reducing data transfer, hence significantly improving energy consumption and latency \cite{3}. 
 Non-volatile memory (NVM) technologies are foundational enablers of this potential paradigm shift. By leveraging a variety of materials systems and switching mechanisms, NVMs are poised to overcome the limitations of traditional architectures and enable scalable, energy-efficient computing platforms \cite{4}. Among NVMs, resistive random-access memory (RRAM) devices particularly stand out as they offer sub-nanosecond latency ($1-10 ns$) at $0.01 pJ/bit$ energy consumption with excellent endurance ($>10^{12}$ cycles) and retention ($>10$ years)\cite{5}. Their intrinsic compatibility with artificial and spiking neural networks makes them key components in the shift beyond conventional von Neumann computing architectures.

%
The resistive switching behavior in RRAM devices is governed by an intricate interplay of physical and chemical mechanisms, including ion migration, phase transitions, and defect-driven filamentary conduction \cite{6}. In filamentary RRAMs, the applied voltage
driven electric field $-$often intensified by Joule heating$-$ induces the redistribution of ions within the switching layer, leading to the formation of conductive filaments that dictate the device’s resistive states and overall performance \cite{7} -\cite{9}. Experimental evidence derived from state-of-the-art characterization techniques highlights that these conductive filaments function as channels that accommodate ionic and electronic transport \cite{10, 11}. A number of parameters, such as the geometry of the device, electrode selection, stoichiometry, and structural characteristics of the resistive switching medium, often originating from thin film deposition conditions \cite{12}, govern the resulting resistive switching dynamics.

Understanding the atomistic origins of filament formation and rupture remains central to unlocking the full potential of RRAMs for next-generation computing. Three dominant mechanisms govern electron and ion transport in these devices: (i) trap states introduced by point defects within the bandgap, (ii) Schottky barrier modulation at the metal/oxide interface, and (iii) the formation and evolution of defect-mediated conduction pathways, i.e., conductive filaments.\cite{13} Bridging atomic-scale defect characteristics—such as formation enthalpy, migration energy, and charge transition levels with materials stoichiometry and device-level performance metrics offers a rational pathway to RRAM optimization. For example, a low oxygen vacancy ($V_o$) formation enthalpy in metal oxide RRAMs may lead to multiple filament formation under applied voltage
driven electric field, causing variability and reduced endurance.\cite{14, 15} 

This study provides atomic-scale insights into the switching mechanisms of RRAM devices, emphasizing the defect-driven processes that define their functionality and reliability. By integrating theoretical modeling with material-level and device-level considerations, we propose a bottom-up framework to guide the rational design of high-performance, energy-efficient memory technologies suited for emerging neuromorphic systems. Our bottom-up simulation-to-experiment co-design framework is demonstrated in Figure \ref{fig1}. 
The rest of the paper proceeds as follows. Section II introduces the methodology for device fabrication and characterization. An extensive set of results is demonstrated in Section III. Section IV concludes the paper by elucidating the contributions and future research outlook.
\begin{figure} [h]
    \centering
    {\label{fig1}\includegraphics[width=0.98\linewidth]{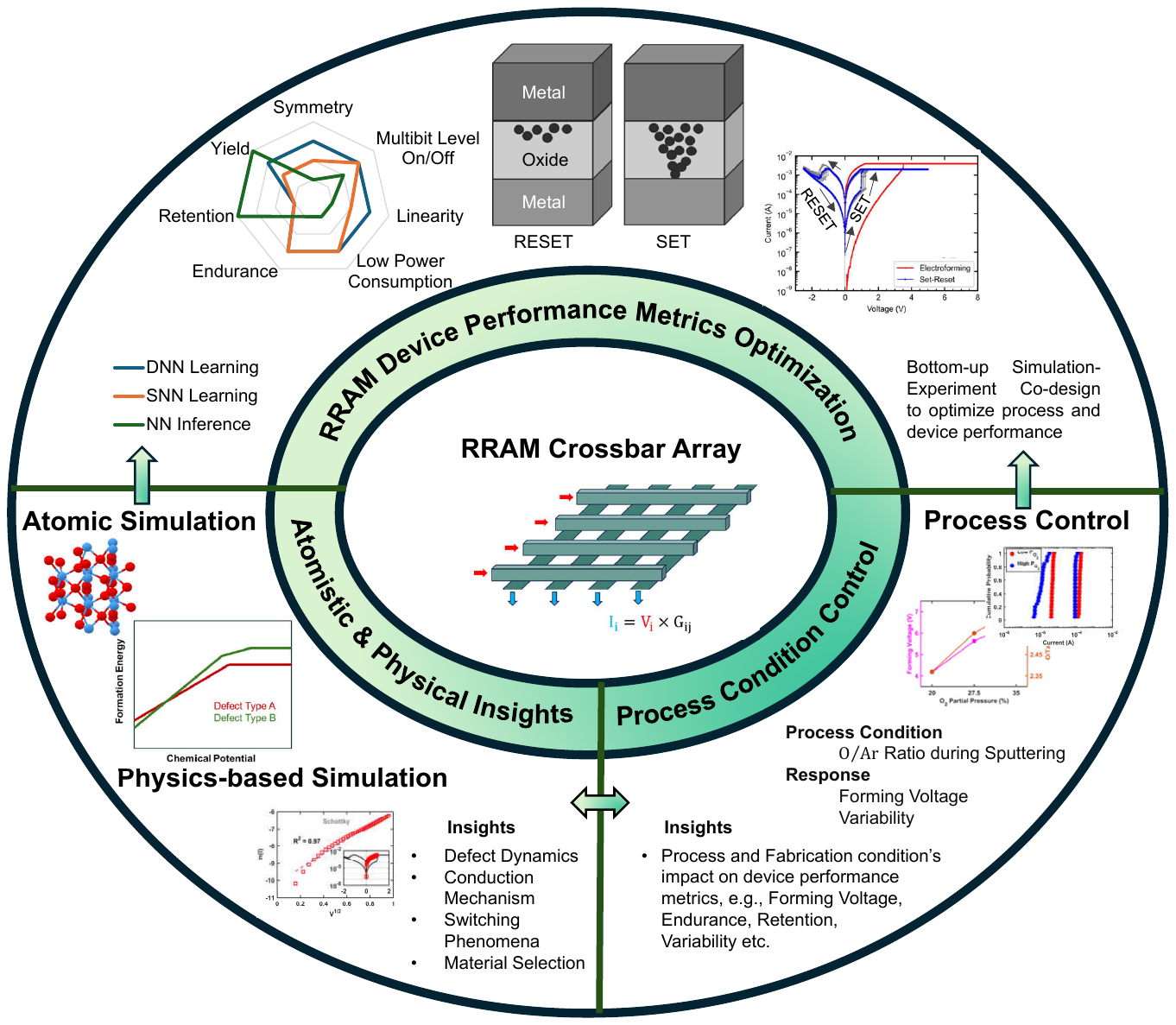}}\hfill
    \caption{Bottom-up simulation-to-experiment co-design framework can help optimize device performance by tuning process control parameters based on atomistic insights}
    \label{fig1}
\vspace{-0mm}
\end{figure}
%

\section{Methodology}
In this study, $Ti/Pt/TaO_{x}/Ta/Pt$ ($1<x<2.5$) crosspoint devices have been used. Fabrication process and device performance metrics are reported in our previous work \cite{16}. The $TaO_{x}$ switching layer was deposited by DC-sputtering. The magnetron sputtering tool used in this process has a variable-gated turbo pump and is capable of maintaining sensitive gas ratios ($<1\%$) for reactive sputtering. It utilizes a DC power source for conductive materials and RF power sources for electrically insulating materials. The system allows control of different parameters, e.g., power intensity (RF or DC), Oxygen/Argon (O/Ar) ratio, substrate temperature, substrate-target distance, and chamber pressure. 
Previous studies have demonstrated the effect of process control parameters, e.g., annealing and deposition conditions, on device metrics, e.g., forming voltage $V_{form}$ \cite{12}. This paper examines two RRAM devices that share the same structural design but differ in the stoichiometry of their switching layers. For this study, $TaO_x$ switching layer is deposited under two different oxygen partial pressures: low, 10\% (2.14 sccm of $O_2$ flow) and high, 30\% (8.25 sccm of $O_2$ flow), and the impact of the oxygen partial pressure ($P_{O_{2}}$) on device performance metrics, particularly ($V_{form}$) and cycle-to-cycle (C2C) variability in low-resistance state (LRS) and high-resistance state (HRS) during SET cycles have been compared. DC resistive switching measurements for this study were performed using a Keysight B1500 system. Current compliance is applied both during electroforming and SET processes to prevent breakdown of the RRAM devices and control filament formation. The conduction mechanisms in the current-voltage (I-V) characteristics have been studied analytically to extract atomistic insights. Finally, these insights are leveraged to establish a connection between process conditions and device performance metrics, providing a framework for performance optimization.

\begin{figure} [h]
    \centering
    \subfloat[]{\label{fig2a}\includegraphics[width=0.49\linewidth]{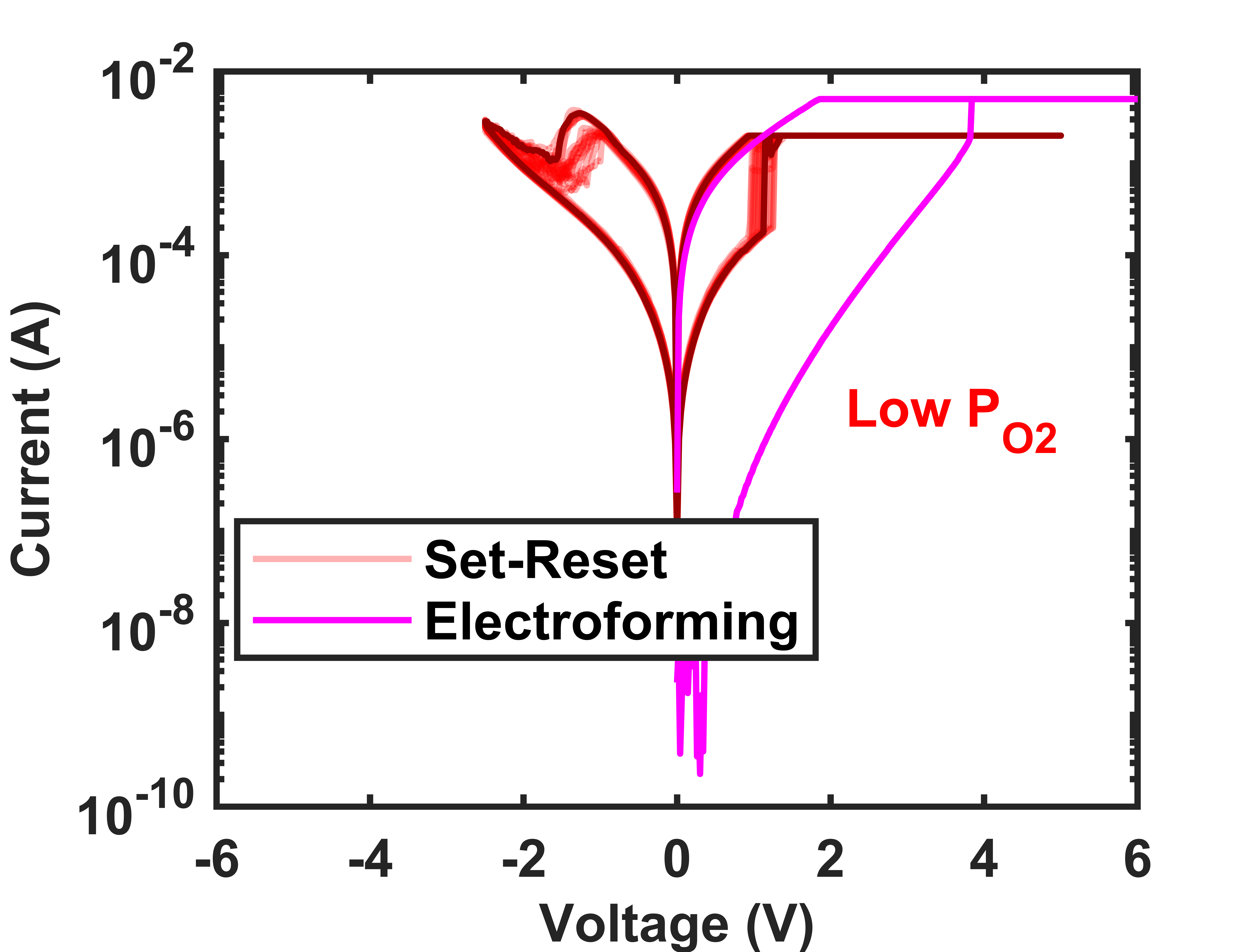}}\hfill
    \subfloat[]{\label{fig2b}\includegraphics[width=0.49\linewidth]{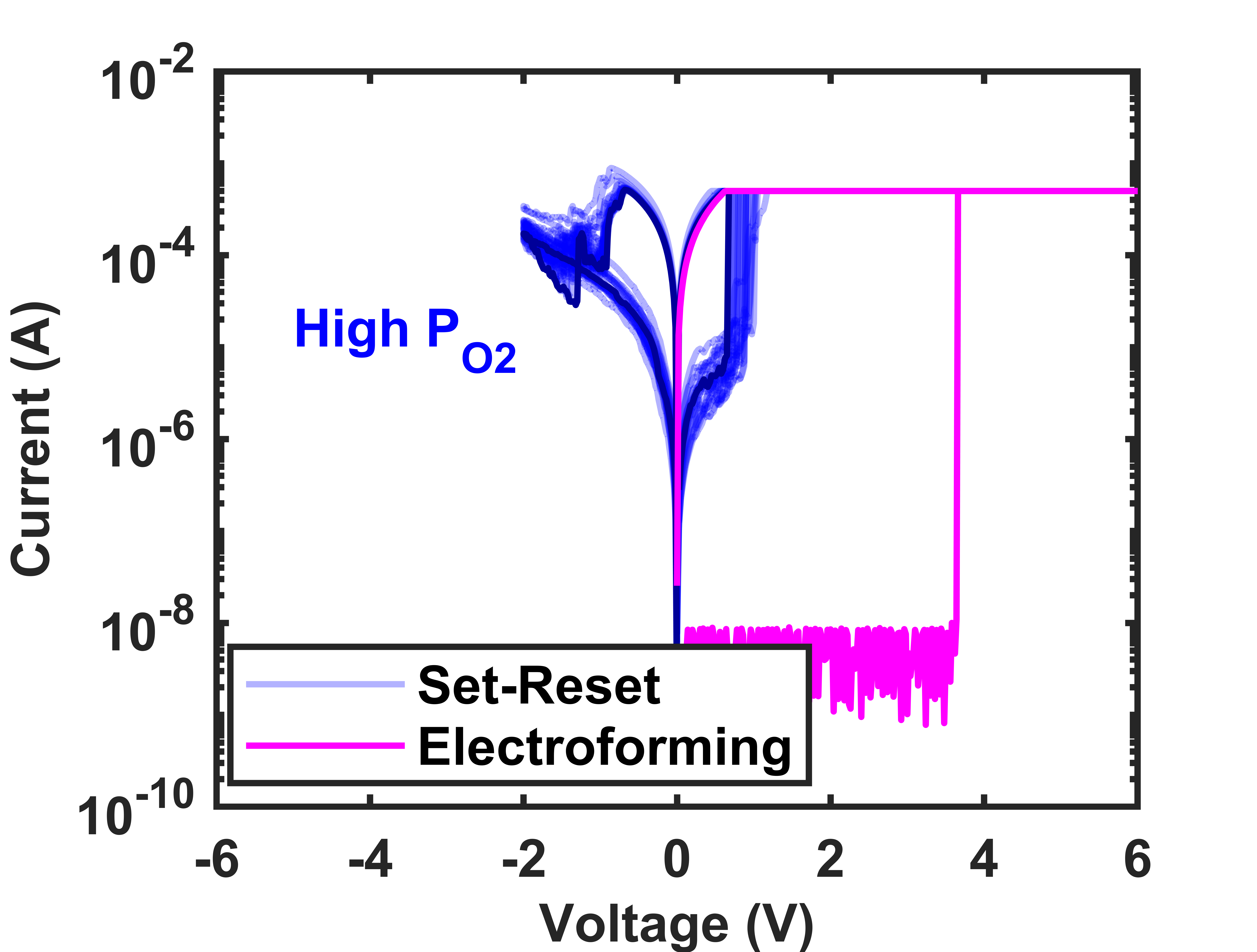}}\\
    \caption{Characteristic I-V plots for devices with $TaO_x$ layer deposited under a) Low O Partial Pressure ($10\%)$ b) High O Partial Pressure ($30\%$)}
    \label{fig2}
\vspace{-0mm}
\end{figure}

\section{Results}
Figure \ref{fig2} presents the characteristic DC I-V plots of two $TaO_x$ RRAM devices demonstrating the electroforming process and 20 cycles of bipolar resistive switching. Initial analysis highlights differences in electroforming dynamics and C2C variability. RRAM device fabricated with a low $P_{O_{2}}$ demonstrates a more gradual electroforming process compared to its high $P_{O_{2}}$ counterpart. C2C variability is also more pronounced in the high $P_{O_{2}}$ RRAM.  
Next, the impact of $P_{O_{2}}$ on $TaO_x$ film stoichiometry and $V_{form}$ has been studied, as shown in Figure \ref{fig3a}. X-ray photoelectron spectroscopy (XPS) has been used to determine the stoichiometry, specifically the O/Ta ratio, for the $TaO_x$ films. $V_{form}$ metric has been extracted from the DC resistive switching measurements of 50 devices for each $P_{O_{2}}$ process condition. Our findings reveal that increasing the oxygen partial pressure during deposition results in more oxygen-rich $TaO_x$ films, characterized by a higher $O/Ta$ atomic ratio. This increased oxygen content correlates with a higher forming voltage, $V_{form}$, indicating a greater initial resistance to filament formation, which is previously reported to be linked to the reduced oxygen vacancy concentration.\cite{18a} 



Figure \ref{fig3b} illustrates the cumulative probability distribution of LRS and HRS currents during the consecutive 20 SET cycles for both oxygen-deficient and oxygen-rich devices, where LRS corresponds to the higher current magnitudes and HRS to the lower ones. The plot reveals that the HRS data for high $P_{O_2}$ exhibits a broader distribution, indicating increased C2C variability. Overall, Fig. 3 demonstrates the impact of a single process control parameter (oxygen partial pressure during DC sputtering) on critical device metrics, e.g., $V_{form}$ and C2C variability. 
\begin{figure}
    \centering
    \subfloat[]{\label{fig3a}\includegraphics[width=.49\linewidth]{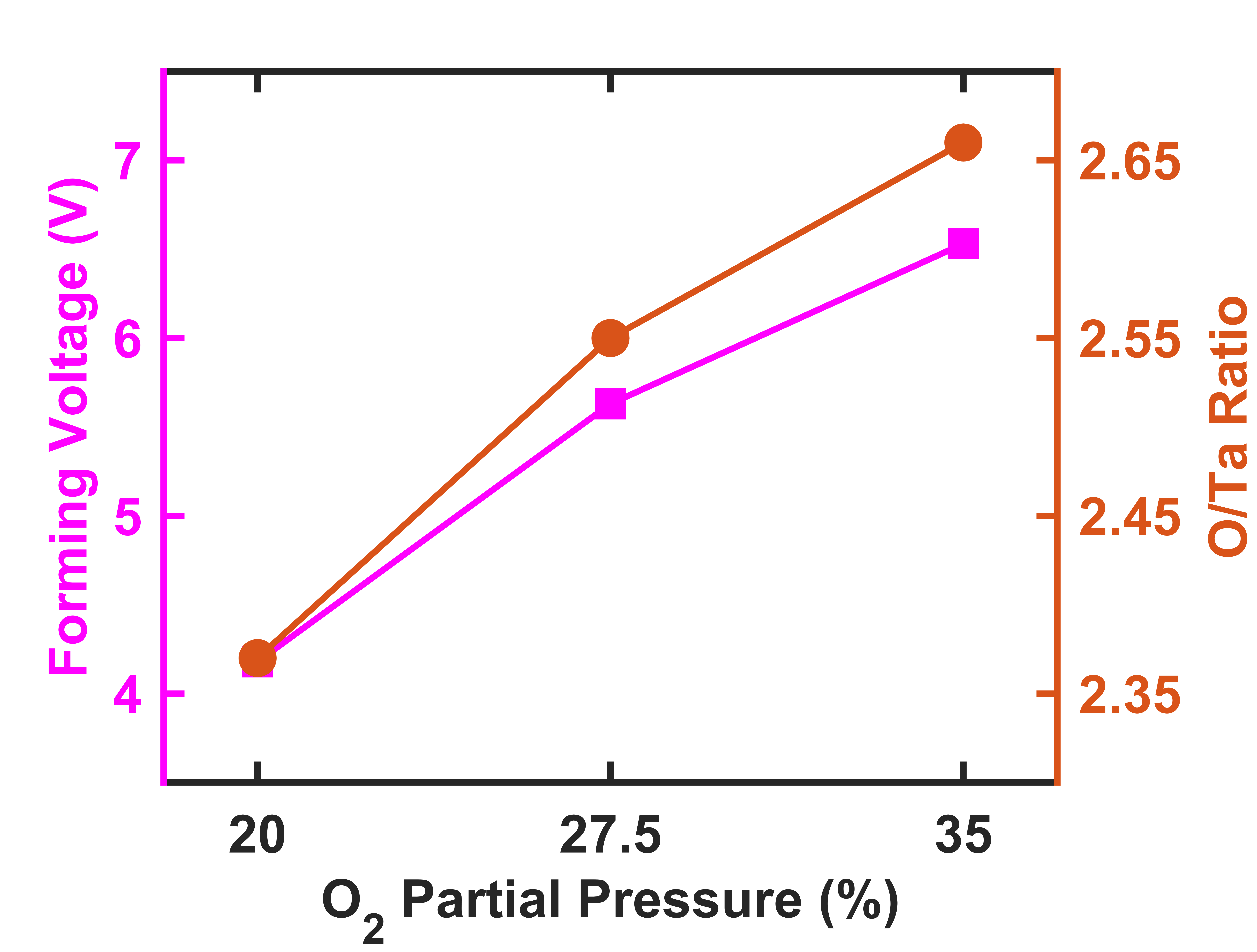}}
    \subfloat[]{\label{fig3b}\includegraphics[width=.49\linewidth]{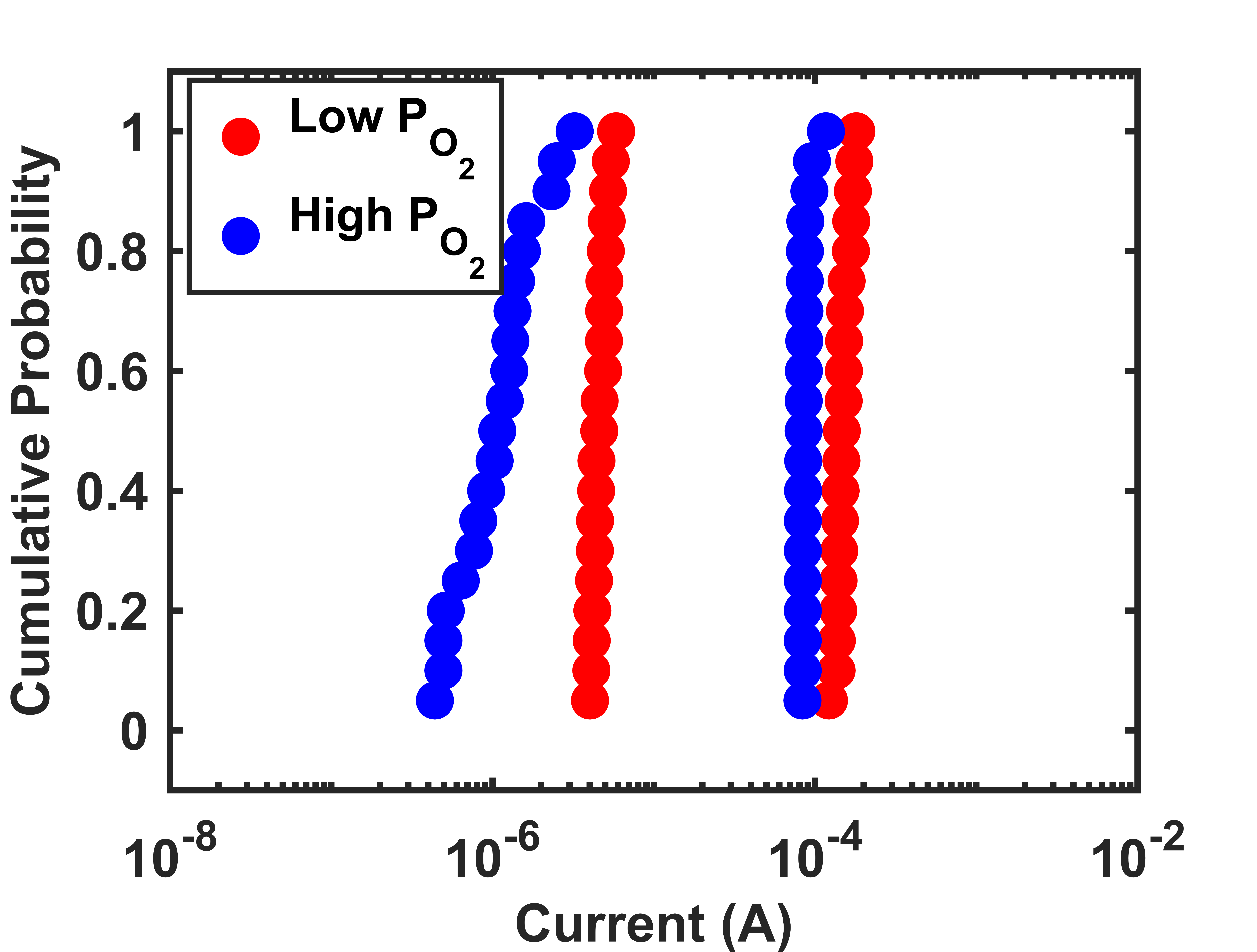}}\hfill
    \caption{Effect of $P_{O_2}$ in a) $V_{form}$ (left) and $TaO_{x}$ stoichiometry (right), b) LRS, HRS data extracted from consecutive SET cycles for both devices}
    \label{fig3}
\end{figure}

Establishing a robust correlation between conduction mechanisms and material properties requires a detailed understanding of the atomistic processes within the switching layer. To this end, we conduct physics-based analytical studies to elucidate the fundamental mechanisms governing device behavior. We begin by examining the HRS during the SET cycle. Figure \ref{fig4}  presents the HRS current behavior of both samples alongside the fitting results for different conduction mechanisms. Fig. \ref{fig4a} shows that the conduction mechanism in HRS current of the switching layer deposited under low $P_{O_{2}}$ is dominated by Poole-Frenkel (PF) ($I\propto{V\times\exp{({\sqrt{V}}})}$) with some deviation below $0.3V$. This phenomenon has previously been studied in $TiN/HfO_x/Pt$ RRAM \cite{17}, which also proposed that interface-limited tunneling, originating from the leaky and defective nature of the oxide layer, accounts for this deviation. In contrast to this behavior, no single conduction mechanism dominated the HRS current in the device with the switching layer deposited under higher $P_{O_{2}}$. In this device, at the lower voltage range ($0-0.3 V$), ohmic conduction ($I\propto{V}$) dominates (Fig. \ref{fig4b}). Then, in the range of $0.3-0.625 V$, the Fowler-Nordheim (FN) conduction mechanism ($I\propto{V^2}\times{\exp{(\frac{1}{V})}}$) shows good alignment (Fig. \ref{fig4c}), indicating the field-dependent electron tunneling more dominant at higher electric fields.  At higher voltages, no single conduction mechanism dominates. Instead, the conduction behavior results from a combination of trap-assisted processes, often influenced by external filamentary contributions within a highly defective switching layer, as observed in previous studies \cite{18}. This observation suggests that a portion of the conductive filament may remain intact after the preceding RESET operation, thereby contributing to the early-stage ohmic behavior observed in the subsequent cycle. Oxygen-rich films with higher $V_O$ defect formation enthalpy contain fewer oxygen vacancies in the pristine stage, which increases the forming voltage but enables more uniform and stable filament formation under strong electric fields. Once formed, metallic-like or continuous filaments can support ohmic transport, consistent with bulk-limited conduction. On the other hand, at higher voltages, electrons can gain sufficient energy to tunnel through the oxide layer due to the localized insulating regions in the filament, showing FN conduction \cite{19}. These distinct behaviors in two different samples indicate that higher oxygen pressure during deposition has introduced higher defect concentration and traps, increasing chances of different types of conduction, resulting in higher variation, which has also been observed in Fig. 2, where the C2C variation is higher in the device where switching layer was deposited under higher $P_{O_{2}}$.
\begin{figure}
    \centering
    \subfloat[]{\label{fig4a}\includegraphics[width=.49\linewidth]{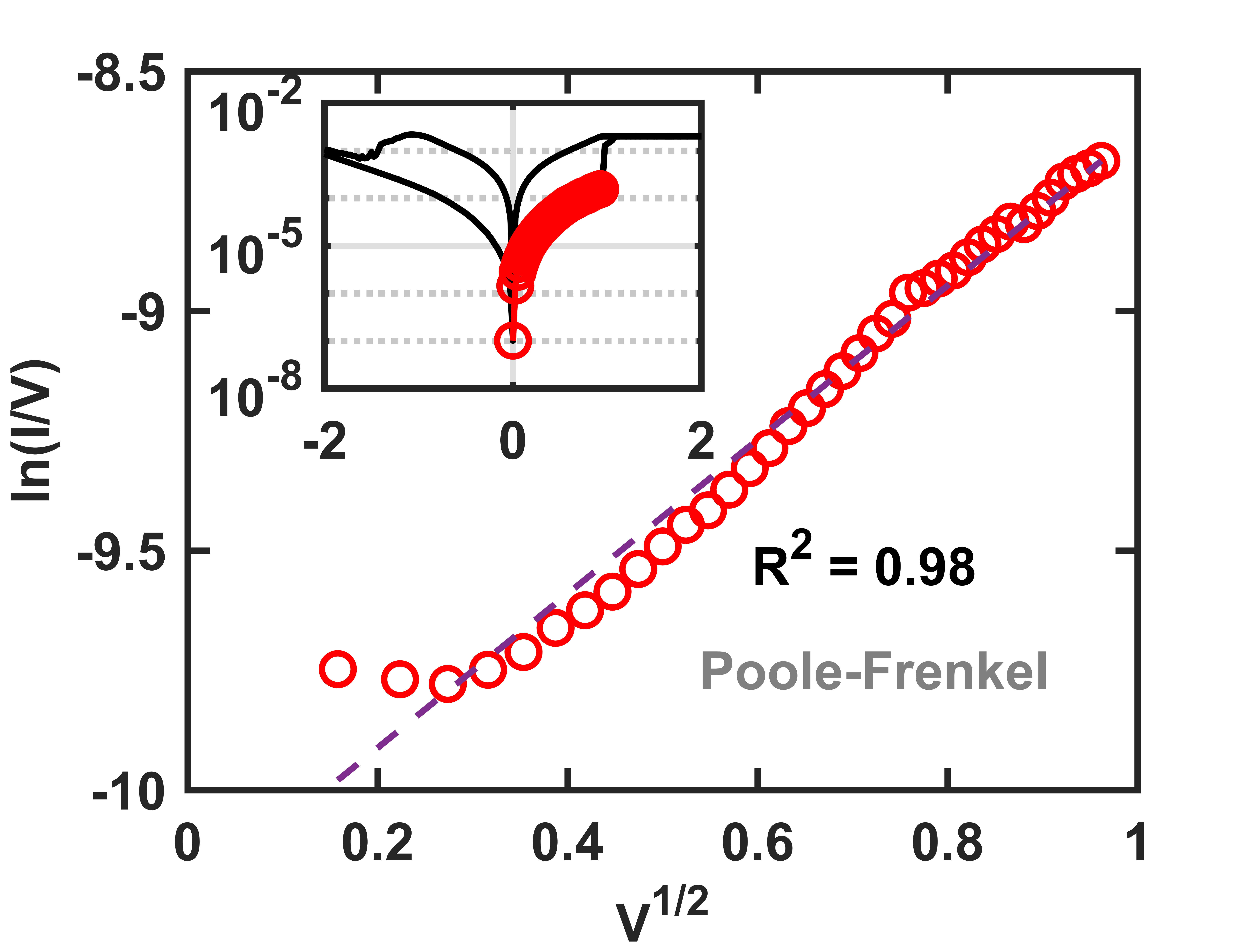}}\hfill
    \\
    \subfloat[]{\label{fig4b}\includegraphics[width=.49\linewidth]{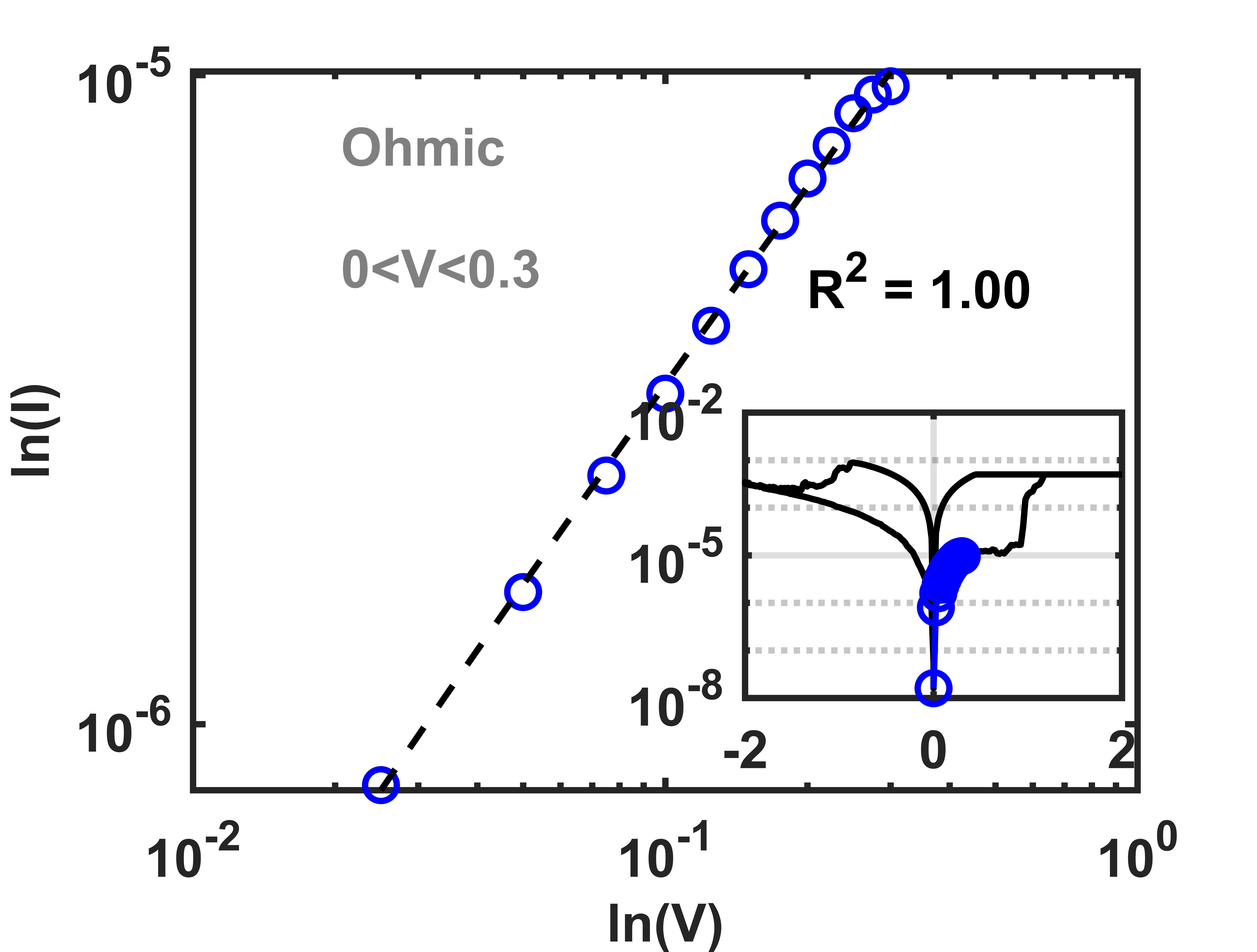}}\hfill
    \subfloat[]{\label{fig4c}\includegraphics[width=.49\linewidth]{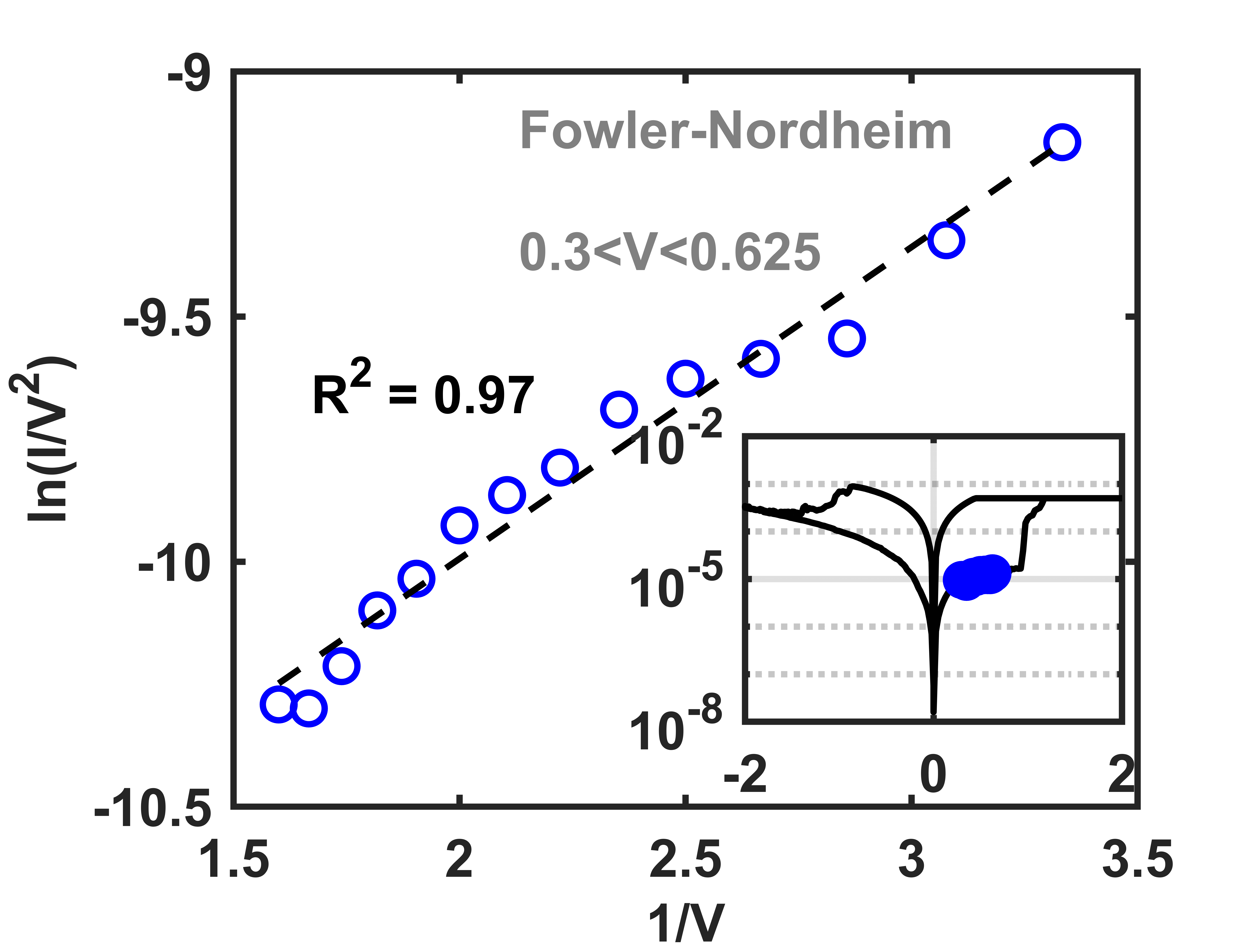}}\hfill
    \caption{HRS Conduction Mechanism in a) Lower Oxygen Pressure Sample and b-c) Higher Oxygen Pressure Sample}
    \label{fig4}
\end{figure}

Lastly, Fig. 5 illustrates the correspondence between the theoretical conduction mechanisms and LRS for both samples. The device with a switching layer deposited under low $P_{O_2}$ showed PF at the higher voltage range, $0.97$ to $1.05V$ (Figure \ref{fig5a}) and Schottky-like conduction ($I\propto{\exp{\sqrt{V}}}$) at lower voltages, $0$ to $\sim 1 V$, (Fig. \ref{fig5b}). However, the device with an oxide layer deposited under high $P_{O_2}$ demonstrated PF at high voltage (Fig. \ref{fig5c}) and ohmic behavior at lower voltages (Fig. \ref{fig5d}). This is expected in these devices as a high electric field generated from the high applied voltage lowers trap-barriers and increases the chance of emission from traps to the conduction band \cite{19}. A high electric field also facilitates the formation of a strong, conductive filament, leading to ohmic conduction at the edge of the SET cycle. In contrast to this filament-dependent conduction, in the low-oxygen sample, the dominance of Schottky conduction suggests that the filament has not properly formed, resulting in Schottky-like interface-limited conduction, further enhanced by the sub-stoichiometric $TaO_x$ layer underneath the electrode. As shown in our previous work \cite{13}, oxygen-deficient $TaO_x$ films yield lower $V_O$ defect formation enthalpies. As the energy cost of vacancy formation is lower for these samples, a dense network of defects can readily align into percolating chains or conductive filaments once an electric field is applied. In this vacancy-rich scenario, the low-resistance state is governed by quasi-metallic or hopping conduction along these filaments, enabling orders-of-magnitude higher current at much lower fields and defining the characteristic filamentary behavior of resistive switching devices.
\begin{figure}
    \centering
    \subfloat[]{\label{fig5a}\includegraphics[width=.49\linewidth]{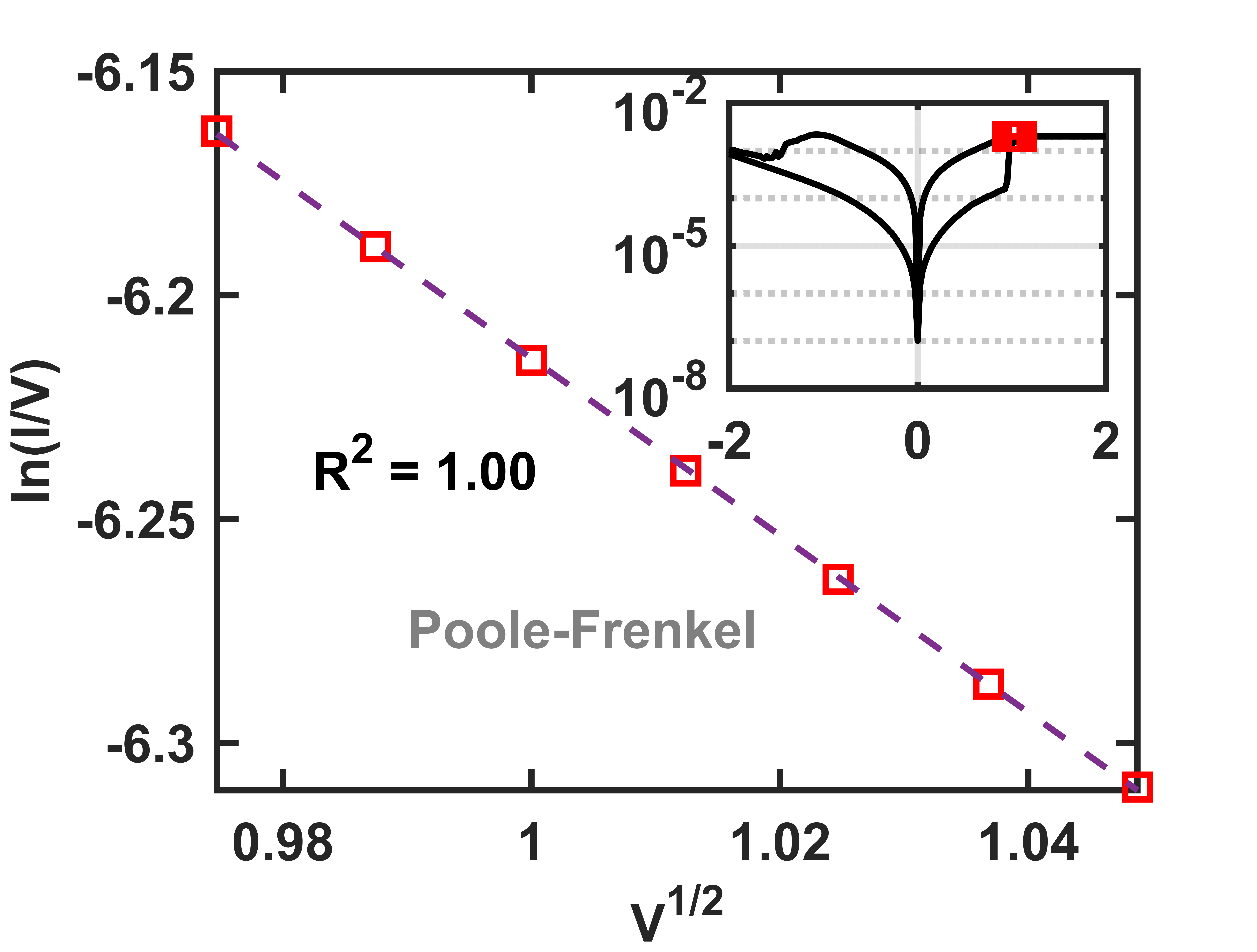}}\hfill
    \subfloat[]{\label{fig5b}\includegraphics[width=.49\linewidth]{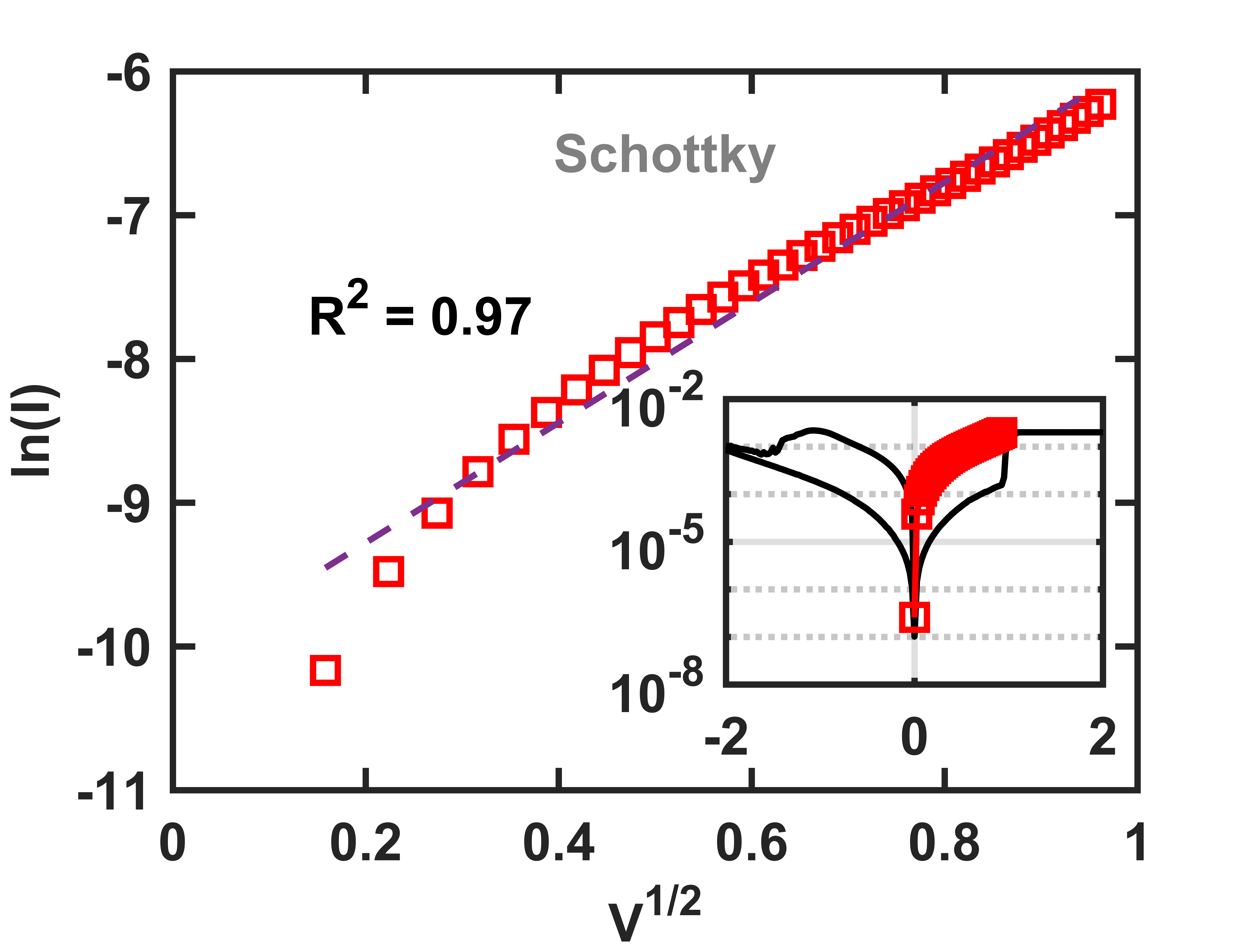}}\hfill
    \\
    \subfloat[]{\label{fig5c}\includegraphics[width=.49\linewidth]{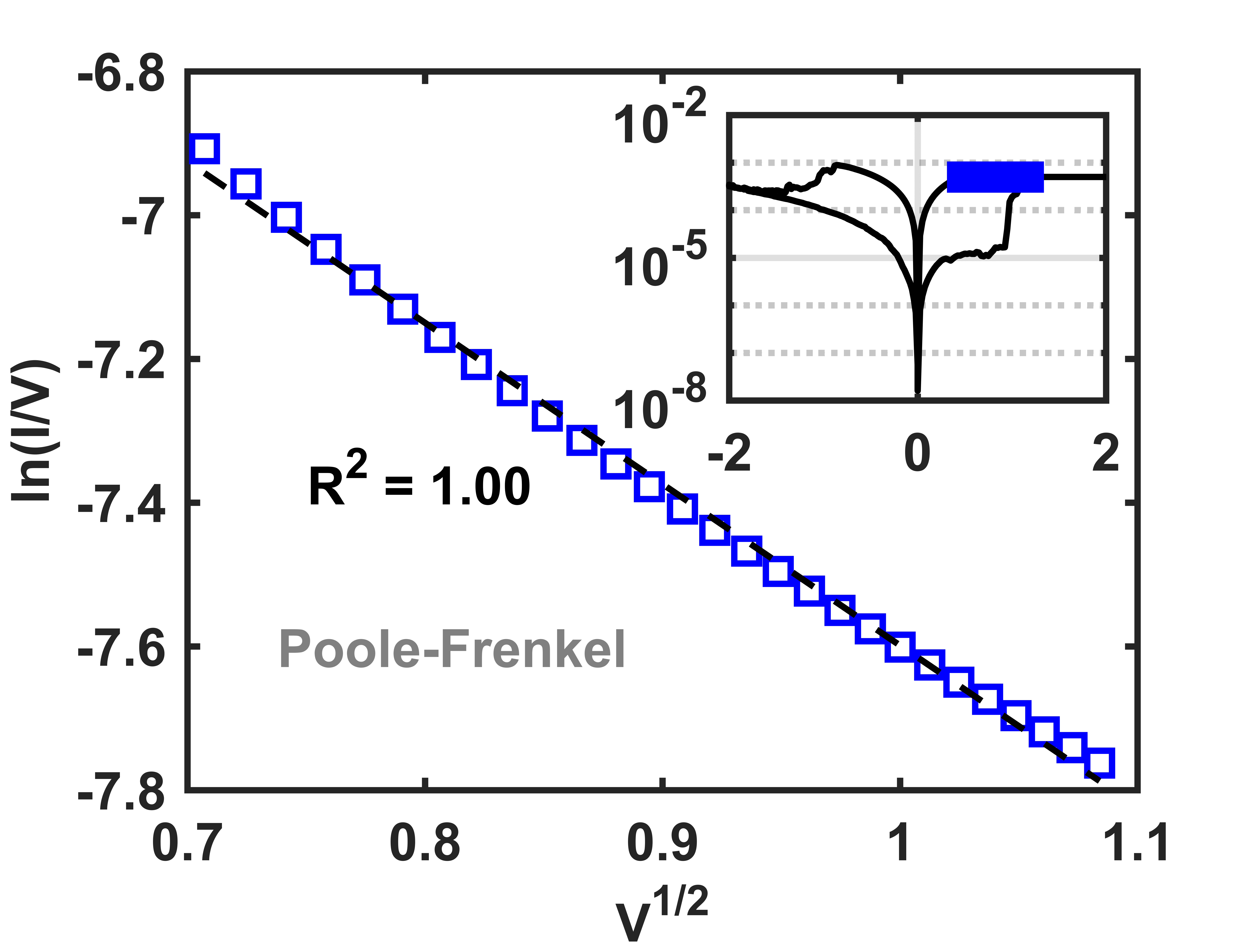}}\hfill
    \subfloat[]{\label{fig5d}\includegraphics[width=.49\linewidth]{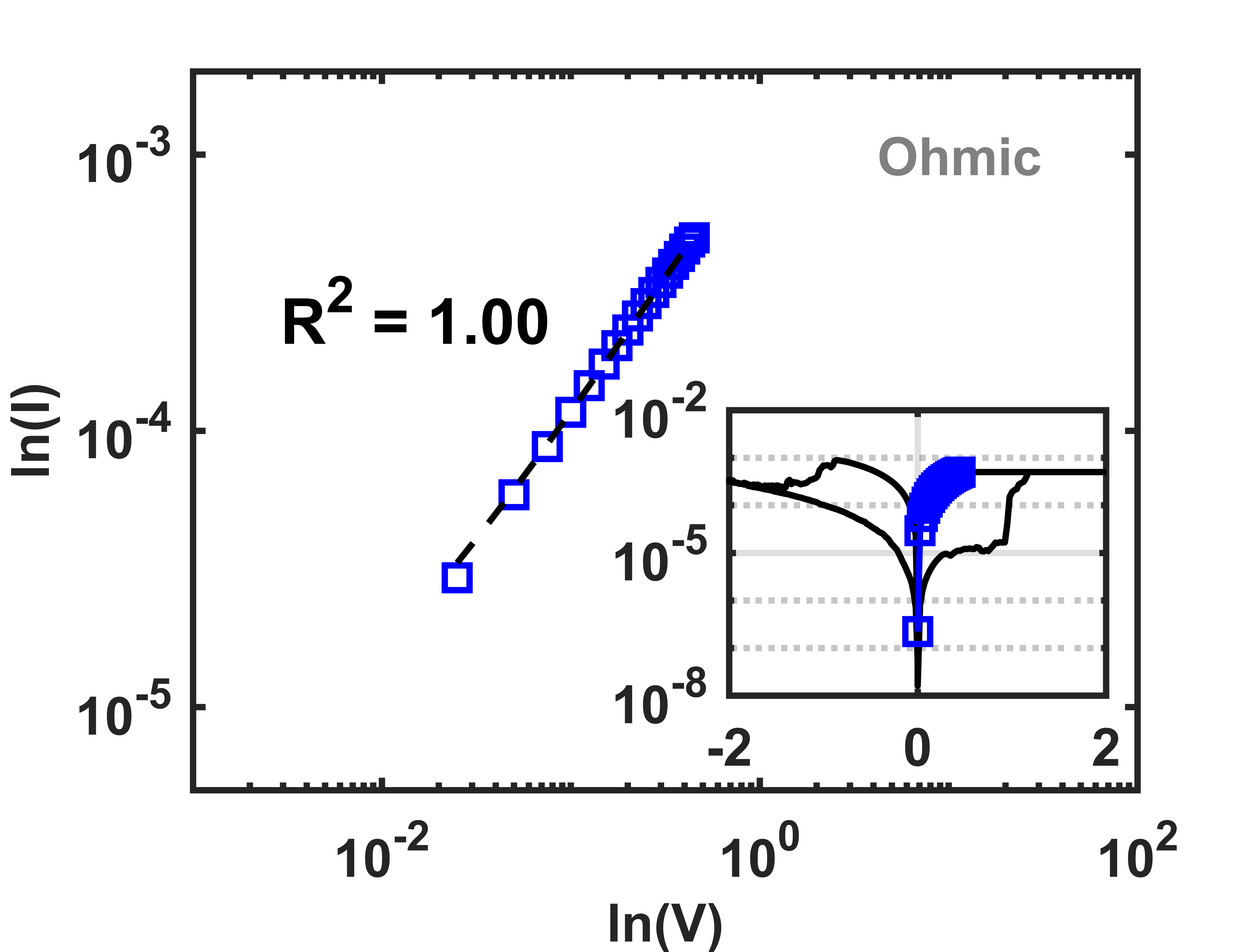}}\hfill
    \caption{LRS Conduction Mechanism in a-b) Lower Oxygen Pressure Sample and c-d) Higher Oxygen Pressure Sample}
    \label{fig3}
\end{figure}

These analytical studies provide a comprehensive atomic-scale analysis of the switching layer in RRAM devices, offering critical insights that can be utilized to refine process control parameters during fabrication. By leveraging these insights, we establish a bottom-up simulation-experiment co-design framework, enabling systematic optimization of device performance. Successful adoption of this bottom-up framework can complement application-specific benchmarking tools such as previously reported MELISO \cite{20}, an end-to-end benchmarking platform for vector-matrix-multiplication (VMM) in RRAM crossbar arrays. MELISO enables the study of the impacts of device properties on VMM error propagation dynamics and their application-specific computational performance metrics. The integration of these two can provide an end-to-end, design technology co-optimization \cite{21} like framework, monumental to overcome stochasticity in process development for successful industry adoption.

\section{Conclusion and Outlook}

This study has emphasized the critical role of atomic-scale mechanisms, particularly defect dynamics and ionic transport, in the governing behavior of the resistive switching of RRAM devices. Through the analysis of conduction mechanisms under varying oxygen stoichiometry conditions, we demonstrate how deposition parameters directly impact filament formation and electronic transport characteristics. 
Specifically, the $TaO_x$ device fabricated under low $P_{O_2} $ exhibited PF conduction at high electric fields and Schottky-like behavior at lower fields, indicating an incomplete or weakly formed filament and interface-limited transport. In contrast, the device deposited under high $P_{O_2} $ showed a transition from PF conduction to ohmic behavior, suggesting the formation of a more continuous and robust filament and a shift toward bulk-limited conduction. These results support the interpretation that high electric fields during the SET process not only promote charge emission from trap states but also drive defect migration and alignment, thereby reinforcing filament connectivity and enabling ohmic transport in the latter stages of switching.


By correlating material processing conditions with electronic transport mechanisms and switching behavior, we reinforce the necessity of atomic-level control in the design and optimization of RRAM devices. The insights presented here pave the way toward a rational, bottom-up design strategy that links defect energetics, stoichiometry, and conduction pathways to scalable device performance—advancing the development of energy-efficient, neuromorphic memory systems.

\section{Acknowledgment}
Gozde Tutuncuoglu would like to acknowledge NSF-CRII grant 2153177 and Michigan Translational Research and Commercialization (MTRAC) Advanced Computing Sprint Award:  Comprehensive Evaluation and Development of Energy Efficient Memristor-Based Systems for Image Recognition Applications.

\vspace{12pt}

\end{document}